\documentclass[a4paper]{article}

\usepackage{INTERSPEECH2022}
\usepackage{makecell}

\title{Automatic Pronunciation Assessment using Self-Supervised Speech Representation Learning}
\name{Eesung Kim, Jae-Jin Jeon, Hyeji Seo, Hoon Kim}

\address{AI Lab, Kakao Enterprise}
\email{\{chris.ekim, jeffrey.j, heize.s, eldon.k\}@kakaoenterprise.com}

\begin{document}

\maketitle
\begin{abstract}
Self-supervised learning (SSL) approaches such as wav2vec 2.0 and HuBERT models have shown promising results in various downstream tasks in the speech community. In particular, speech representations learned by SSL models have been shown to be effective for encoding various speech-related characteristics. In this context, we propose a novel automatic pronunciation assessment method based on SSL models. First, the proposed method fine-tunes the pre-trained SSL models with connectionist temporal classification to adapt the English pronunciation of English-as-a-second-language (ESL) learners in a data environment. Then, the layer-wise contextual representations are extracted from all across the transformer layers of the SSL models. Finally, the automatic pronunciation score is estimated using bidirectional long short-term memory with the layer-wise contextual representations and the corresponding text. We show that the proposed SSL model-based methods outperform the baselines, in terms of the Pearson correlation coefficient, on datasets of Korean ESL learner children and Speechocean762. Furthermore, we analyze how different representations of transformer layers in the SSL model affect the performance of the pronunciation assessment task. 
\end{abstract}
\noindent\textbf{Index Terms}: automatic pronunciation assessment, pronunciation scoring, self-supervised speech representation learning, wav2vec 2.0, HuBERT

\section{Introduction}
The need for English-as-a-second-language (ESL) learners to improve their English pronunciation is increasing owing to globalization. The computer-assisted pronunciation training (CAPT) system, which can conduct assessments and provide detailed feedback on pronunciation proficiency, is thus attracting attention as an ESL learning service and platform \cite{rogerson2021computer, kyriakopoulos2021deep}. There are two technical approaches to the CAPT system: mispronunciation detection and diagnosis (MDD) \cite{ zechner2009automatic,leung2019cnn, yan2020end,lin2021attention,lo2020effective,feng2020sed,lo2021improving,peng21e_interspeech, wu21h_interspeech,xu21k_interspeech} and automatic pronunciation assessment \cite{lin2021attention,li2011context,witt2000phone,yu2015using,chen2018end,Kyriakopoulos2018,peng21e_interspeech}. MDD is a task of detecting pronunciation errors by calculating multiple measures using estimated and canonical phones from an automatic speech recognizer. In this study, we focus on the automatic pronunciation-scoring problem, which estimates a pronunciation score based on pronunciation-relevant characteristics from spoken English data to achieve a high correlation with the scores annotated by experts.

Most previous studies that measured speech pronunciation used automatic speech recognition (ASR) systems and acoustic features to estimate the pronunciation score. With the ASR system, SpeechRater \cite{zechner2009automatic} is an effective method for predicting pronunciation using several handcrafted speech features. Another common feature used for pronunciation assessment is the Goodness of Pronunciation (GOP) measure \cite{witt2000phone,sudhakara2019improved} and its variations, while using the ASR system and forced alignment. In addition, many studies have utilized acoustic features, including prosody, intensity, rhythm, and cepstrum, for pronunciation assessments.

Several recent studies on automatic pronunciation assessment were based on deep neural networks \cite{lin2021attention,yu2015using, chen2018end,Kyriakopoulos2018,lin21j_interspeech}. Bidirectional long short-term memory (BLSTM)-based time-sequence features and time-aggregated features were optimized using a multilayer perceptron (MLP) for automatic scoring \cite{yu2015using}. BLSTM, with attention structure, simultaneously estimates the pronunciation score with both acoustic and linguistic cues \cite{chen2018end}. In \cite{Kyriakopoulos2018}, phone distance metric features extracted from a Siamese network based on BLSTM were used for end-to-end training using an attention mechanism. In \cite{lin21j_interspeech}, deep features from the acoustic model and the self-attention mechanism for a scoring module were utilized directly. According to end-to-end (E2E) ASR advances, E2E-based research has been introduced to pronunciation assessment with competitive results. An E2E model consisting of two encoders for text and audio followed the attention mechanism to combine the two features \cite{lin2021attention}.

Recently, self-supervised learning (SSL) \cite{liu2020mockingjay, liu2021tera, baevski2020wav2vec, hsu21_interspeech, hsu2021hubert} has shown promising results in the downstream tasks of speech processing applications, such as ASR, phoneme recognition, emotion recognition, speaker diarization, and speaker verification \cite{baevski2020wav2vec,hsu2021hubert,yang2021superb}. These studies applied contextual representations using pre-trained models or a fine-tuned model for related tasks. In particular, speech representations learned across the transformer layers of SSL models have been shown to be highly effective in learning high-level representations by encoding various speech-related properties and linguistic information contents \cite{ma2021probing,pasad2021layer,shah2021all}. However, for the CAPT system, researchers have investigated the capability of the wav2vec 2.0 model only in MDD tasks \cite{peng21e_interspeech, wu21h_interspeech,xu21k_interspeech}.

\begin{figure*}[t]
  \centering
  \includegraphics[width=0.32\textheight]{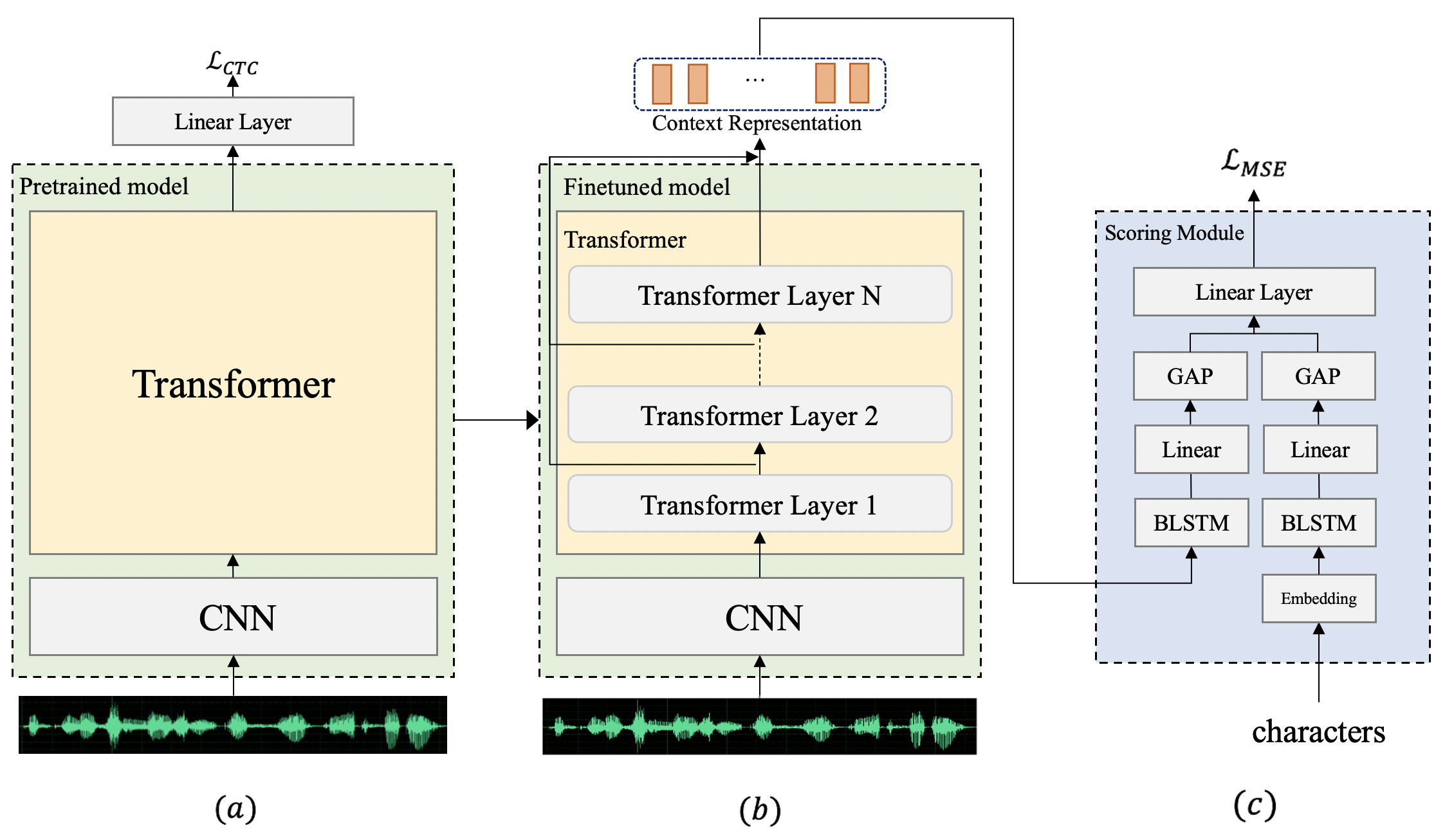}
  \caption{Overall procedure of the proposed method for automatic pronunciation assessment based on SSL models.}
\end{figure*}

In this paper, we proposed an SSL-model-based automatic pronunciation assessment method. We hypothesized that SSL would be effective for learning pronunciation-relevant latent representations. To test this hypothesis, we propose a novel pronunciation assessment method that incorporates pre-trained SSL models to fine-tune and adapt them to the pronunciation of non-native English language learners. Subsequently, the weighted average of all context representations was extracted across the transformer layers of the SSL models. Finally, BLSTM was constructed on top of the models incorporating script information to estimate the utterance-level pronunciation score. We observed a significant improvement from fined-tuned SSL models over other baselines in terms of the Pearson correlation coefficient (PCC) with two datasets of Korean children: ESL learners (KESL) and Chinese speakers (Speechocean762). To the best of our knowledge, this is the first study on pronunciation scoring using the SSL method.

\section{Method} \label{sec:method}
An overview of the proposed method is shown in Figure 1. The method includes three stages: (i) fine-tuning the pre-trained contextual transformer to adapt to a non-native spoken data environment with connectionist temporal classification (CTC) loss \cite{graves2006connectionist}, (ii) extracting layer-wise contextual representations from the contextual transformer, and (iii) estimating the utterance-level pronunciation score using the neural network constructed on top of the contextual transformer. Our target SSL-based architecture is the wav2vec 2.0 and HuBERT models.

\subsection{Pre-trained Models}
Wav2vec 2.0 \cite{baevski2020wav2vec}, which has been well verified \cite{yang2021superb} is an effective structure for learning representations from speech data using SSL models. The wav2vec 2.0 model consists of a convolutional encoder, context encoder, and quantization module. The convolutional encoder contains time-stacking convolutional layer maps that transform the raw waveform into latent speech representations. The context encoder, consisting of multiple transformer blocks, takes a latent speech representation and outputs the context representation. The latent speech representations are quantized into an embedding with a fixed number of latent representations stored in a codebook for the prediction task. The contrastive score between context representation and vector-quantized embedding is maximized during network training \cite{oord2018representation}.

HuBERT \cite{hsu2021hubert} has been investigated for outperforming wav2vec 2.0 for multiple downstream recognition and generation tasks. It uses the same structure as wav2vec 2.0, which consists of a convolutional encoder followed by a transformer context encoder. In the training phase, HuBERT builds pseudo-labels through iterative refinement of clustering with mel-frequency cepstrum coefficients (MFCC) and hidden units of transformers, while training wav2vec 2.0 through a quantization module using Gumbel-softmax \cite{baevski2020wav2vec}. With the discovered hidden units, the model was trained using cross-entropy loss over the masked regions only.

\subsection{Fine-tuning}
Fine-tuning in SSL has the advantage of adapting to the characteristics of a small dataset. Following the wav2vec 2.0 and HuBERT fine-tuning procedure \cite{baevski2020wav2vec,hsu2021hubert}, pre-trained models are fine-tuned with the given non-native spoken training data, which are optimized with the CTC loss $L_{CTC}$ \cite{graves2006connectionist}. We assumed that it would be more beneficial for the model to adapt to the non-native spoken dataset environment with the ASR fine-tuning process for pronunciation scoring. The final hidden state of the transformers is convoluted using a 1D convolutional layer and then fed into a softmax layer, as shown in Figure 1 (a). All model weights, except for the convolutional layer, were fine-tuned.

\subsection{Pronunciation Representation Modeling}

Recent studies \cite{pasad2021layer, shah2021all} have shown that various properties including acoustic and linguistic information tend to be encoded in different layers of transformers. We hypothesized that the transformers in the SSL models would learn pronunciation-related features when the model is optimized using a pronunciation dataset. To capture all encoded information related to pronunciation effectively, we extract layer-wise context representations by averaging the hidden states of the context representations of all transformer layers, as illustrated in Figure 1 (b). In addition, we investigated the effectiveness of utilizing the convolution layer output and context representation outputs of different transformer layers in the SSL models on the pronunciation assessment task.

\subsection{Scoring Module for Pronunciation Assessment}
To achieve the final goal of predicting pronunciation scores, this study was approached as a regression problem by optimizing the mean square error loss $L_{MSE}$ between the predicted and human-annotated scores. As shown in Figure 1 (c), we adopted the BLSTM model to reflect the dynamics of the acoustic and linguistic representations for the utterance-level score, as in \cite{yu2015using}\cite{chen2018end}. First, we generated two unidirectional encoded vectors in the forward and backward directions from BLSTM with the audio context representations of the transformer layers of the SSL models. Then, we obtained the utterance-level full audio context vector by concatenating the two unidirectional encoded vectors, followed by the linear layer. To encode the dynamics of linguistic information for the spoken utterance, characters of the script of the spoken utterance were converted into embedded vectors by the embedding layer, and then the embedded vectors were transformed to two BLSTM-based unidirectional encoded vectors followed by a linear layer to obtain the bidirectional full linguistic context vector. The final score was obtained by applying global average pooling (GAP) of audio context vectors and linguistic context vectors over time dimension and script characters, respectively, and then a linear layer.

\section{EXPERIMENT}
\label{sec:pagestyle}

\subsection{Dataset}
For the assessment of non-native English spoken pronunciation, two different human-labeled datasets were utilized to demonstrate the effectiveness of our proposed method. The first corpus was an in-house dataset recorded by Korean ESL learner (KESL) children. It consisted of 17800 utterances by 300 Korean speakers with ages ranging from 10 to 12 years. Five native experts annotated the sentence level using five pronunciation continuous measures ranging from 1 to 5, including the holistic impression of pronunciation, segmental accuracy, stress, pauses, and intonation. Utterance-level scores were obtained by averaging the scores of five experts for each label. A holistic score distribution of the dataset is depicted in Figure 2. The second dataset was a public dataset called Speechocean762 \cite{zhang2021speechocean762}. It contained 5000 English sentences recorded by 250 English non-native English speakers, in which the gender and age of the speakers were proportionately balanced. The dataset provides multidimensional scores, such as accuracy, completeness, fluency, and prosody, in terms of word-level, phoneme level, and sentence level. Sentence-level labels were used to evaluate pronunciation. The dataset was divided into a training set and test set at a ratio of 5:5. The sampling rate of all the speech data was 16,000 Hz.

\subsection{Baselines}
For traditional features for pronunciation-scoring tasks, we used time-aggregated features, time-sequence handcrafted acoustic features, GOP, and their combination features, which were used in previous studies \cite{yu2015using, chen2018end, lin21j_interspeech}. For \textbf{\emph{GOP}}, the frame-level posterior was generated using the DNN-based acoustic model as the likelihood ratio between the forced alignment likelihood and the maximum likelihood obtained from the ASR engine \cite{hu2015improved}. The frame-level posterior matrix was generated by forwarding propagation on the native acoustic model, and the matrix was used for forced alignment and computing to obtain the GOP-based features, whose definitions can be found in \cite{hu2015improved}. The ASR system used in this study was an in-house DNN-HMM hybrid ASR system trained on 4000 h of native and non-native spoken English data. In addition, we utilized the time-aggregated features (\textbf{\emph{AggFeat}}) used in previous works \cite{yu2015using,chen2018end} based on SpeechRater \cite{zechner2009automatic} related to several aspects of the speech construct, including fluency, rhythm, intonation, stress, pronunciation, grammar, and vocabulary use. For the time-sequence features indicated as \textbf{\emph{SeqFeat}}, the mean and standard deviation of 23 low-level descriptors using the eGeMAPS set \cite{eyben2015geneva}, including the MFCC, loudness, pitch, jitter, and shimmer over the segment implemented in OpenSmile \cite{eyben2010opensmile} were extracted. Feature-wise zero-mean and unit-variance normalizations were used for all features. For the scoring module of the baselines, we averaged the segment features along the time dimension using GAP and then concatenated the baseline features, followed by two linear layers, 256 and 1 unit, respectively, to predict the pronunciation score.

\begin{figure}[t]
  \centering
  \includegraphics[height=0.12\textheight]{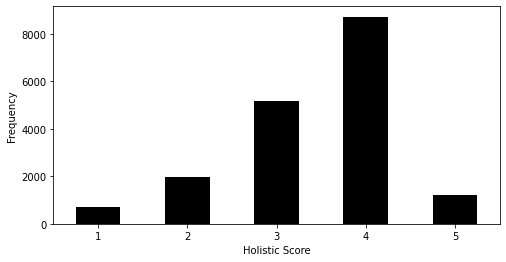}
  \caption{The distribution of holistic score of Korean ESL children dataset annotated by experts.}
\end{figure}

\subsection{Experimental Setup}

Experiments on SSL models used the pre-trained wav2vec2-large-960h \cite{baevski2020wav2vec}, wav2vec2-base-960h \cite{baevski2020wav2vec}, wav2vec2-large-robust \cite{baevski2020wav2vec}, HuBERT-base-ls960h \cite{hsu2021hubert}, and HuBERT-large-ls960h \cite{hsu2021hubert} from Fairseq \cite{ott-etal-2019-fairseq}. For the fine-tuning stage of all models, we finetuned 150k steps with eight batch sizes and used an Adam optimizer, where the learning rate was warmed up from 1e-4 with a 1k warm-up step using the training implementation on \emph{HuggingFace} \cite{wolf-etal-2020-transformers}. The model with the lowest word error rate score was used for the development set. To train the pre-trained model with CTC, the vocabulary was built from all distinct letters of the training and test data, including \emph{CTC blank}, \emph{apostrophe}, and \emph{space} tokens. A 10-fold cross-validation method based on the speaker was used. For all scoring modules, we experimented with a single-layer BLSTM with 128 hidden and 64 embedding dimensions. The model was trained using the Adam optimizer with a learning rate of 1e-4 and early stopping with a patience of 3 on the validation loss. To evaluate the models, we compared the PCC of the predicted and human-annotated scores.

\subsection{Results}

\begin{table} [ht]
% \toprule
\centering
    \setcounter{table}{0}
	\begin{tabular}{c|c|c|c}
	\Xhline{4\arrayrulewidth}

	\multicolumn{1}{l}{{\textbf{Method}}}&
	\multicolumn{1}{|c}{\textbf{KESL}}&
	\multicolumn{2}{|c}{\textbf{Speechocean762}} \\ \cline{2-4}

	\multicolumn{1}{p{2.5cm}}{}&
	\multicolumn{1}{|p{1.0cm}}{\centering\textbf{Holistic}}& 
	\multicolumn{1}{|p{1.0cm}}{\centering\textbf{Fluency}}&
	\multicolumn{1}{|p{1.0cm}}{\centering\textbf{Prosodic}}\\

    \Xhline{3\arrayrulewidth}
	\multicolumn{1}{l}{\text{GOP}}&
	\multicolumn{1}{|c}{0.63}&
	\multicolumn{1}{|c}{0.65}&
	\multicolumn{1}{|c}{0.64}\\

	\multicolumn{1}{l}{\text{Agg + Seq}}&
	\multicolumn{1}{|c}{0.55}&
	\multicolumn{1}{|c}{0.51}&
	\multicolumn{1}{|c}{0.59}\\
	
	\multicolumn{1}{l}{\text{Agg + Seq + GOP}} &
	\multicolumn{1}{|c}{0.64}&
	\multicolumn{1}{|c}{0.67}&
	\multicolumn{1}{|c}{0.66}\\
	
    \Xhline{3\arrayrulewidth}
	\multicolumn{1}{l}{pre-trained} &
	\multicolumn{1}{|c}{}&
	\multicolumn{1}{|c}{}&
	\multicolumn{1}{|c}{}\\ 
    % \hspace*{-1.8em}
    
	\multicolumn{1}{l}{\quad \text{wav2vec2 Base}} &
	\multicolumn{1}{|c}{0.65}&
	\multicolumn{1}{|c}{0.72}&
	\multicolumn{1}{|c}{0.72}\\

	\multicolumn{1}{l}{\quad \text{wav2vec2 Large}} &
	\multicolumn{1}{|c}{0.71}&
	\multicolumn{1}{|c}{0.72}&
	\multicolumn{1}{|c}{0.72}\\

	\multicolumn{1}{l}{\quad \text{wav2vec2 Robust}} &
	\multicolumn{1}{|c}{0.76}&
	\multicolumn{1}{|c}{0.73}&
	\multicolumn{1}{|c}{0.73}\\

	\multicolumn{1}{l}{\quad \text{HuBERT Base}} &
	\multicolumn{1}{|c}{0.69}&
	\multicolumn{1}{|c}{0.72}&
	\multicolumn{1}{|c}{0.71}\\

	\multicolumn{1}{l}{\quad \text{HuBERT Large}} &
	\multicolumn{1}{|c}{0.75}&
	
	\multicolumn{1}{|c}{0.75}&
	\multicolumn{1}{|c}{0.74}\\

    \Xhline{3\arrayrulewidth}
	\multicolumn{1}{l}{Finetuned} &
	\multicolumn{1}{|c}{}&
	\multicolumn{1}{|c}{}&
	\multicolumn{1}{|c}{}\\
	
	\multicolumn{1}{l}{\quad \text{wav2vec2 Base}} &
	\multicolumn{1}{|c}{0.68}&
	\multicolumn{1}{|c}{0.73}&
	\multicolumn{1}{|c}{0.72}\\

	\multicolumn{1}{l}{\quad \text{wav2vec2 Large}} &
	\multicolumn{1}{|c}{0.78}&
	\multicolumn{1}{|c}{0.73}&
	\multicolumn{1}{|c}{0.72}\\

	\multicolumn{1}{l}{\quad \text{wav2vec2 Robust}} &
	\multicolumn{1}{|c}{0.79}&
	\multicolumn{1}{|c}{0.75}&
	\multicolumn{1}{|c}{0.74}\\

	\multicolumn{1}{l}{\quad \text{HuBERT Base}} &
	\multicolumn{1}{|c}{0.75}&
	\multicolumn{1}{|c}{0.74}&
	\multicolumn{1}{|c}{0.73}\\

	\multicolumn{1}{l}{\quad \text{HuBERT Large}} &
	\multicolumn{1}{|c}{\textbf{0.82}}&
	\multicolumn{1}{|c}{\textbf{0.78}}&
	\multicolumn{1}{|c}{\textbf{0.77}}\\

    \Xhline{5\arrayrulewidth}
% \bottomrule
\end{tabular}
\caption{Performance of proposed method in terms of PCC with different approaches, based on the English pronunciation dataset including speech from Korean children and the Speechocean762 dataset.}

\label{tab:1}
\end{table}

To evaluate the performance of the SSL-based methods, we explored improvements to the SSL-based pronunciation assessment models with two corpora: the KESL and Speechocean762 datasets. Table 1 compares the performance of SSL-based pre-trained models, fine-tuned models of wav2vec 2.0 (wav2vec2), and HuBERT with traditional baselines. All fine-tuned models were fine-tuned on two non-native datasets, KESL and Speechocean72. All SSL models used layer-wise context representations. First, both HuBERT and wav2vec2 exhibit performance improvements over the existing baseline model by using only the layer-wise context representations of the pre-trained model, as shown in Table 1. In particular, the wav2vec2 robust pre-trained model \cite{hsu21_interspeech} showed high performance, and it can be inferred that wav2vec2 robust is a pre-trained model with speech in real scenarios, making the representation more robust for non-native speaker conditions. Second, we observed that the fine-tuned models consistently outperformed the pre-trained models in both the wav2vec2 and HuBERT models. This shows that the fine-tuned model for ASR gains is beneficial for estimating pronunciation scores compared to the pre-trained model, which did not adapt to the non-native spoken data environment. Finally, we can see that the fine-tuned HuBERT Large model achieved the best results for PCCs of holistic, fluency, and prosodic measures on the KESL and Speechocean762 datasets. We can see that the holistic, fluency, and prosodic aspects of the HuBERT Large model outperform those of the wav2vec2 robust model for the PCCs by $0.03$, $0.03$, and $0.03$, respectively.

\subsection{Ablation Studies}

In this section, we focus on our experiment using the HuBERT model as an example to study the factors affecting the performance of the SSL model in automatic pronunciation assessment.

\subsubsection{Effect of contextual representations of the HuBERT model}
First, we explore the effectiveness of using representations from different transformer layers of the pre-trained and fine-tuned HuBERT large models for the pronunciation assessment task. As shown in Figure 3, we can observe that there is a performance difference among the different layers, and the tendencies of the PCC according to different layers are similar among the prosodic and SpeechOcean762 datasets. Figure 3 shows that the performance of both the fine-tuned and pre-trained models in the upper layer parts (from the 11th to 22nd layers) performed significantly better than those in the lower layer parts (from 1st to 9th layer). This indicates that the HuBERT Large model can better learn pronunciation-related information in the upper parts. Table 2 shows that transformer-based contextual representations perform better than convolutional encoder representations. We can observe that the hidden layer from the 20th layer obtains the best PCC among the transformer layers. Finally, the layer-wise context vector, which is the average of the hidden states of all layers, achieved the best PCC score. This indicates that it is beneficial for the pronunciation assessment task to use the information of all layers, including the various properties of speech.

\begin{figure}[t]
  \centering
  \includegraphics[width=0.32\textheight]{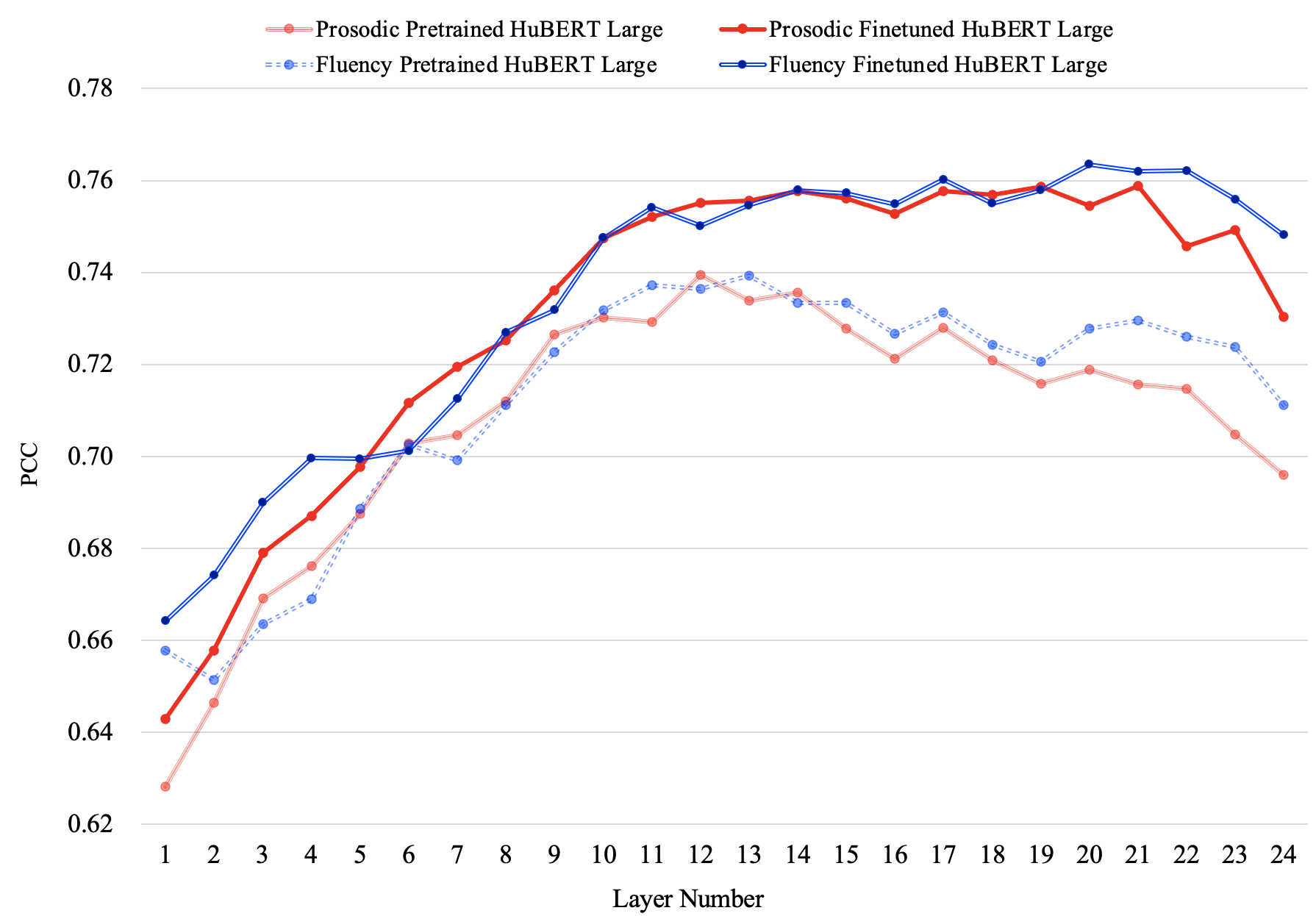}
  \caption{PCCs when using the hidden states of different transformer layers of the fine-tuned HuBERT Large model as input to the BLSTM scoring module.}
\end{figure}

\begin{table} [ht]
  \begin{center}
	\begin{tabular}{p{1.5cm}|p{1.0cm}|p{1.0cm}|p{1.0cm}}
	\Xhline{5\arrayrulewidth}
	\multicolumn{1}{c}{{\textbf{\footnotesize Feature}}}&
	\multicolumn{1}{|c}{\textbf{\footnotesize KESL}}&
	\multicolumn{2}{|c}{\textbf{\footnotesize Speechocean762}} \\ \cline{2-4}

	\multicolumn{1}{p{1.5cm}}{ }&
	\multicolumn{1}{|p{1.0cm}}{\centering\textbf{{\footnotesize Holistic}} }&
	\multicolumn{1}{|p{1.0cm}}{\centering\textbf{{\footnotesize Fluency}} }&
	\multicolumn{1}{|p{1.0cm}}{\centering\textbf{{\footnotesize Prosodic}}} \\
	
	\Xhline{3\arrayrulewidth}
	\multicolumn{1}{c}{Local} &
	\multicolumn{1}{|c}{0.56} &
	\multicolumn{1}{|c}{0.60} &
	\multicolumn{1}{|c}{0.62} \\

	\multicolumn{1}{c}{\text{Layer 20}} &
	\multicolumn{1}{|c}{0.81} &
	\multicolumn{1}{|c}{0.76} &
	\multicolumn{1}{|c}{0.76} \\

	\multicolumn{1}{c}{\text{All Layers (Proposed)}} &
	\multicolumn{1}{|c}{\textbf{0.82}} &
	\multicolumn{1}{|c}{\textbf{0.78}} &
	\multicolumn{1}{|c}{\textbf{0.77}} \\

	\Xhline{5\arrayrulewidth}
    \end{tabular}
  \end{center}
  
    \caption{Comparison of the performance of local representation of a convolutional layer, the contextual representation, and layer-wise contextual representation of the transformer layers in HuBERT Large model.}
  
  \label{tab:3}
\end{table}

\subsubsection{Comparison of different regression methods}

The last ablation involved a pronunciation-scoring module. We tested commonly used regression methods for pronunciation task, including linear regression (LR) and the MLP on top of the transformer of HuBERT to aggregate an utterance-level signal vector. The PCCs of holistic, fluency, and prosodic of the BLSTM scoring module outperformed all the baseline performances on two corpora, KESL and Speechocean762, by 0.02, 0.05, and 0.03, respectively. The results demonstrate that the BLSTM-based scoring module performs better than simple regression methods with average pooling method.

\begin{table} [ht]
  \begin{center}
	\begin{tabular}{p{1.8cm}|p{1.0cm}|p{1.0cm}|p{1.0cm}}
	\Xhline{5\arrayrulewidth}
	\multicolumn{1}{c}{{\textbf{\footnotesize Scoring Module}}}&
	\multicolumn{1}{|c}{\textbf{\footnotesize KESL}}&
	\multicolumn{2}{|c}{\textbf{\footnotesize Speechocean762}} \\ \cline{2-4}

	\multicolumn{1}{p{1.8cm}}{ }&
	\multicolumn{1}{|p{1.0cm}}{\centering\textbf{{\footnotesize Holistic}} }&
	\multicolumn{1}{|p{1.0cm}}{\centering\textbf{{\footnotesize Fluency}} }&
	\multicolumn{1}{|p{1.0cm}}{\centering\textbf{{\footnotesize Prosodic}}} \\
	\Xhline{3\arrayrulewidth}

	\multicolumn{1}{c}{\text{LR}} &
	\multicolumn{1}{|c}{0.78} &
	\multicolumn{1}{|c}{0.72} &
	\multicolumn{1}{|c}{0.72} \\
	
	\multicolumn{1}{c}{\text{MLP}} &
	\multicolumn{1}{|c}{0.80} &
	\multicolumn{1}{|c}{0.73} &
	\multicolumn{1}{|c}{0.74} \\

	\multicolumn{1}{c}{\text{BLSTM} (Proposed)} &
	\multicolumn{1}{|c}{\textbf{0.82}} &
	\multicolumn{1}{|c}{\textbf{0.78}} &
	\multicolumn{1}{|c}{\textbf{0.77}} \\

	\Xhline{5\arrayrulewidth}
    \end{tabular}
  \end{center}

    \caption{Comparison of the performance applying the different types of scoring modules on top of layer-wise transformer layers of finetuned HuBERT Large model.}
  \label{tab:4}
\end{table}

\section{CONCLUSION}
\label{sec:conclusion}
In this paper, we presented an automatic pronunciation assessment method utilizing effective contextual representations of SSL models such as wav2vec 2.0 and HuBERT. We showed that pre-trained SSL models are beneficial for estimating pronunciation-scoring tasks. In addition, fine-tuning SSL models with CTC is beneficial for improving pronunciation-scoring performance. Finally, we demonstrated the effectiveness of the layer-wise representation of transformer layers from the perspective of a pronunciation assessment task. The experiments were conducted with two datasets, KESL and Speechocean762, in terms of PCC. In future research, we plan to study the effectiveness of fine-tuning SSL models in terms of pronunciation score.

\bibliographystyle{IEEEtran}

\bibliography{Bibliography}

% \begin{thebibliography}

% \end{thebibliography}

\end{document}